\documentclass[apj]{emulateapj}
\usepackage{graphicx}

\usepackage{color}

\shorttitle{Supermassive Seeds for Supermassive Black Holes}
\shortauthors{Johnson, Whalen, Li and Holz}

\begin{document}

\title{Supermassive Seeds for Supermassive Black Holes}
\author{Jarrett L. Johnson\altaffilmark{1}, Daniel
  J. Whalen\altaffilmark{1,2}, Hui Li\altaffilmark{1} and Daniel
  E. Holz\altaffilmark{3}}

\affil{$^{1}$Nuclear and Particle Physics, Astrophysics and Cosmology
  Group (T-2), \\
Los Alamos 
National Laboratory, Los Alamos, NM 87545; jlj@lanl.gov}
\affil{$^{2}$McWilliams Fellow, Department of Physics, Carnegie Mellon University, Pittsburgh, PA 
15213}
\affil{$^{3}$Enrico Fermi Institute, Department of Physics, and Kavli
  Institute for Cosmological Physics, \\
University of Chicago, Chicago, IL  60637}

\begin{abstract}
Recent observations of quasars powered by
supermassive black holes (SMBHs) out to $z\ga 7$ constrain
both the initial seed masses and the growth of the most 
massive black holes (BHs) in the early universe. 
Here we elucidate the implications of the radiative feedback from early generations
of stars and from BH accretion for popular models for the formation and
growth of seed BHs.
We show that by properly accounting for (1) the limited role of mergers in growing seed 
BHs as inferred from cosmological simulations of early star formation and radiative feedback, (2) the
sub-Eddington accretion rates of BHs expected at the earliest times, and (3) the
large radiative efficiencies $\epsilon$ of the most massive BHs inferred from observations
of active galactic nuclei at high redshift ($\epsilon$ $\ga$ 0.1), we are led to the conclusion that the
initial BH seeds may have been as massive as $\ga 10^5 M_{\odot}$.
This presents a strong challenge to the Population III seed model,
which calls for seed masses of $\sim$ 100 $M_{\odot}$ and, even with constant Eddington-limited accretion, 
requires $\epsilon$ $\la$ 0.09 to explain the highest-$z$ SMBHs in 
today's standard $\Lambda$CDM cosmological model.  It is, however,  
consistent with the prediction of the direct collapse scenario of SMBH
seed formation, in which a supermassive primordial star forms in a 
region of the universe with a high molecule-dissociating
background radiation field, and collapses directly into a $10^4$--$10^6$
$M_{\odot}$ seed BH.  These results corroborate recent
cosmological simulations and observational campaigns which suggest that these massive BHs were 
the seeds of a large fraction of the SMBHs residing in the
centers of galaxies today.
 \end{abstract}


\section{Introduction}
What is the origin of the SMBHs that power the most
distant quasars?  There are currently three main scenarios
for the formation of the initial ``seed'' BHs from which these SMBHs
grew (e.g. Haiman 2009, 2012; Natarajan 2011; Volonteri 2010, 2012): (1) massive
Population (Pop)~III seeds, which form from the collapse of $\sim
30$--$300\,M_{\odot}$ 
primordial stars in dark matter (DM) minihalos with total masses of $\sim
10^5$--$10^6 M_{\odot}$ at redshifts $z\ga20$ (e.g. Madau \& Rees 2001); (2)
supermassive stellar remnant seeds, which form with initial masses of
$10^4$--$10^6M_{\odot}$ from the direct collapse of $\simeq$10$^4$ K primordial
gas in atomic cooling DM halos with total masses of $\sim$10$^7$--10$^8$
$M_{\odot}$ at $z$ $\ga$10 (e.g. Bromm \& Loeb 2003); (3) and seeds formed from
the collapse of $\sim$10$^3$ $M_{\odot}$ stars created in runaway collisions in
dense stellar clusters at $z$ $\sim$ 10--20 (e.g. Devecchi \& Volonteri
2009; Davies et al. 2011). 

Distinguishing between these three scenarios poses significant
challenges, in large part because the high-$z$ regime in which these
seed BHs are born is too distant to be probed directly by
existing facilities.  However, significant progress has been made in
detecting $\ga 10^9 M_{\odot}$ SMBHs powering
quasars at $z\ga 6$, the existence of which provides significant
constraints on the nature and growth of their BH seeds (e.g. Willott
et al. 2003; Fan 2006).  The strongest
constraints to date come from the $\simeq 2 \times 10^9\,M_{\odot}$
SMBH inferred to be powering a quasar at $z = 7.085$
(Mortlock et al. 2011).   Given the $\la 800$ Myr available for the
growth of such a massive BH, only the most optimistic models can
explain their origin from $\sim$100 $M_{\odot}$ Pop~III seeds,
suggesting that more massive seeds may have a role to play
(e.g. Baumgarte \& Shapiro 1999; Tyler et al. 2003; Shapiro 2005; Volonteri \& Rees 2006; Tanaka
et al. 2012).  

Also consistent with supermassive seeds are
observations of high-$z$ quasars powered by SMBHs that are very
massive compared to the stellar component of their host galaxies
(e.g. Wang et al. 2010, 2013; Willott et al. 2013), much
more so than
would be predicted following the observed relation in the local
universe (e.g. Ferrarese \& Merritt 2000; Gebhart et al. 2000; but see
van den Bosch et al. 2012 for a notable
exception).\footnote{Interestingly, there is also some observational
  evidence that local BHs may have been seeded by direct collapse (Greene 2012).}
As discussed by Agarwal et al. (2013), such relatively massive BHs in early
galaxies are easily accommodated in the SMS seed model. 

At the same time, a growing body of theoretical work 
is providing renewed support for the long-standing supermassive stellar remnant model
(e.g. Fowler \& Hoyle 1964; Appenzeller \& Fricke 1972;
Shapiro \& Teukolsky 1979; Bond et al. 1984).
In particular, there are now strong suggestions that the conditions
required for the formation of supermassive stars (SMSs) may be satisfied
more often in the early universe than previously assumed (Wise et
al. 2008; Regan \& Haehnelt 2009; Sethi et al. 2010; Shang et al. 2010; Bellovary et
al. 2011; Wolcott-Green et
al. 2011; Agarwal et al. 2012; Inayoshi \& Omukai 2012; Johnson et
al. 2012a; Latif et al. 2012, 2013; Petri et al. 2012; Choi et
al. 2013; Prieto et al. 2013).  In addition, modeling the
evolution of rapidly accreting SMSs (e.g. Begelman 2010; Hosokawa et
al. 2012; Inayoshi et al. 2013)
and the radiative feedback they exert during their growth
(Johnson et al. 2012b; but see also Dotan \& Shaviv 2012) shows that they
can attain masses $\ga$ 10$^5$ $M_{\odot}$ before collapsing to BHs.
These massive BH seeds could in principle grow to SMBHs much more quickly than
10--10$^3$ $M_{\odot}$ seeds.  This is especially true in light of numerous
other recent theoretical (e.g. Volonteri et al. 2005; Pelupessy et al. 2007; Alvarez et al. 2009; Milosavljevi{\' c}
et al. 2009a; Noble et al. 2009; Park \& Ricotti 2011, 2012a,b; Jeon et al. 2012) and observational
(e.g. Elvis et al. 2002; Wang et al. 2006, 2009; Davis \& Laor 2011; Bambi et al. 2012; Li et al. 2012) results which suggest that accretion onto BHs in the
early universe was suppressed due to radiative feedback.

Here we examine the limits that can be placed on the initial seed
masses and growth of the most massive {high-$z$} BHs, by accounting
for the radiative feedback from early stellar generations and from the
accretion of gas onto these BHs.  In the next
Section we show that adopting appropriate
initial Pop~III seed DM halo masses and accounting for the limited 
production of Pop~III seeds due to the Lyman-Werner background
radiation field allows us to conclude that mergers play only a small role 
in the growth of the most massive BHs at high-$z$.  
In Section 3 we survey the possible range of time-averaged growth
rates and radiative efficiencies of high-$z$ SMBHs in the context of growth 
solely via gas accretion.  In Section 4 we discuss how
local radiative feedback from both accreting BHs and their progenitor stars 
acts to limit the rate of growth of Pop~III seed BHs and, in particular,
hinders the growth of the least massive BHs.  Finally, we discuss the 
implications for the main models of seed formation in Section 5, and
we summarize our results in Section 6.

\section{Limited Growth via Mergers due to Global Lyman-Werner Feedback}
Here we make a simple, but novel, argument for why the role of mergers in growing the most massive
high-$z$ BHs must be limited, in part, due to global
molecule-dissociating, Lyman-Werner radiative feedback.
This will allow us to dramatically simplify our discussion of SMBH
growth in the early universe in later Sections, by focusing solely on growth via gas accretion.

Many previous works have examined the possibility of SMBHs growing from Pop~III
seeds via mergers and gas accretion (Menou et al. 2001; Haehnelt 2003; Islam et
al. 2003; Yoo \&
Miralda-Escud{\' e} 2004; Volonteri et al. 2003, 2005; Tanaka \&
Haiman 2009).  If sufficient mass can be locked up in the seeds,
merging the BHs is an effective means of growing a single SMBH
that avoids the radiative feedback limitations associated with accreting gas. 
Most recently, Tanaka et al. (2012) have reported that mergers of
Pop~III remnant seeds born in DM halos with virial temperatures
$T_{\rm vir}$ = 400 K enhance the final masses of SMBHs by a factor of
10--100.\footnote{These authors also account for the reduced rate of
  gas accretion onto BHs when they are kicked out into the low
  density regions of their host halos due to the emission of
  gravitational waves during mergers (e.g. Favata et al. 2004;
  Merritt et al. 2004; see also Madau et al. 2004).  They also
  emphasize that the bulk of the growth in mass comes from gas
  accretion, not mergers.}  
However, accounting for the substantially more massive halos, with $T_{\rm
  vir}$ $\sim$ 1000--2000 K,\footnote{The virial temperature of a
  halo increases with halo mass as $T_{\rm vir}$ $\propto$ $M_{\rm
    halo}^{\frac{2}{3}}$ (e.g. Barkana \& Loeb 2001).} in which Pop~III star formation is found
to occur in cosmological simulations (e.g. Yoshida et al. 2003; O'Shea \& Norman 2007), 
even at the earliest epochs (e.g. Gao et al. 2007), implies a decrease
in the host halo abundance by a factor of 1--2 orders of magnitude
(as found from large-scale cosmological simulations of hierarchical structure
formation; e.g. Reed et al. 2007).  
Assuming the same star formation efficiency
and Pop~III initial mass function from Tanaka et al. (2012) this, in turn,
suggests that the number of mergers that occur in the assembly of a
high-$z$ SMBH is 1--2 orders of magnitude lower than Tanaka et al. find,
which implies that mergers are likely to contribute only minimally to the growth
of SMBHs at high-$z$ (see also Madau et al. 2004).\footnote{Consistent with this are the results of recent large-scale cosmological 
simulations tracking the build-up of SMBHs at high
redshift which suggest that only a negligible fraction of their mass is
acquired in mergers (DeGraf et al. 2012).}  This dramatically
limits the masses to which BHs can grow from Pop~III seeds, in particular.

The rate of mergers that grow SMBHs from Pop~III seeds is
further reduced due to the build-up of the molecule-dissociating
Lyman-Werner (LW) radiation field which acts to slow the collapse of
primordial gas in DM halos, and thus lowers the Pop~III star formation rate
(SFR) (e.g. Haiman et al. 1997; Glover \& Brand 2001; Machacek et al. 2001;
Ricotti et al. 2001; Ciardi \& Ferrara 2005; Mesinger et al. 2006;
Wise \& Abel 2007; O'Shea \& Norman 2008).   
The most recent cosmological simulations tracking the build-up
of the LW radiation field suggest that the Pop~III SFR is reduced by a
factor of a few compared to the SFR in the absence of LW feedback
(Ahn et al. 2012; Hummel et al. 2012; Johnson et al. 2012a; Wise et al. 2012).\footnote{Tanaka et al. (2012) also considered the impact of LW feedback
and found it to be insignificant; however, this is likely due, at least in part, 
to their use of the fitting formula 
from Machacek et al. (2001) which has been shown to significantly underestimate the minimum 
halo mass for star formation in the presence of LW radiation (see e.g. O'Shea \& Norman 2008).}
Accounting for this further reduction in the number density of Pop~III seed
BHs, beyond the reduction due to the more massive host halos, we can
conclude that mergers are likely to be responsible for only a
small portion of the growth of SMBHs. 
The vast majority of the growth must
be due instead to accretion of gas (see also, e.g., Hopkins \& Quataert 2010)
and perhaps to a much lesser extent DM (e.g. Hu et al. 2006, Guzm{\' a}n
\&  Lora-Clavijo 2011). 
This justifies our simplified approach in later Sections of focusing on
the growth of BHs solely from accretion and neglecting mergers.  
In turn, as we will show in the next Sections, the limited role of
mergers in early SMBH growth implies a strong challenge to the Pop~III
seed model, which relies on frequent mergers to grow the most massive BHs.

A lower rate of (gas poor) mergers of seed BHs also suggests higher 
values for the spins of the seeds (e.g. Gammie et al. 2004; Volonteri et al. 2005, 2012; Berti
\& Volonteri 2008), since mergers of BHs with randomly oriented spins
will tend to spin down fast rotating BHs (e.g. Hughes \& Blandford 2003). 
 This translates into a
higher radiative efficiency of accretion, since more energy can be
extracted from the spin of the BH (Blandford \& Znajek 1977) and from 
the hotter accretion disk that extends further inward towards
the horizon of a faster-spinning BH
(e.g. Novikov \& Thorne 1973).\footnote{It is worth noting that the
  spins of local SMBHs are found to be $\ga$ 60 percent of the
  maximum allowed value (Brenneman et al. 2011), which suggests that
  their radiative efficiencies are higher than the value expected for a
  non-rotating BH, $\epsilon$ $\ga$ 0.06 (e.g. Noble et al. 2011).}
That we find mergers to be relatively unimportant is then broadly
consistent with the inferred high radiative efficiencies of high-$z$
SMBHs, which we shall discuss in Section 4.

\begin{figure}
  \begin{center}
    \leavevmode
      \epsfxsize=8.5cm\epsfbox{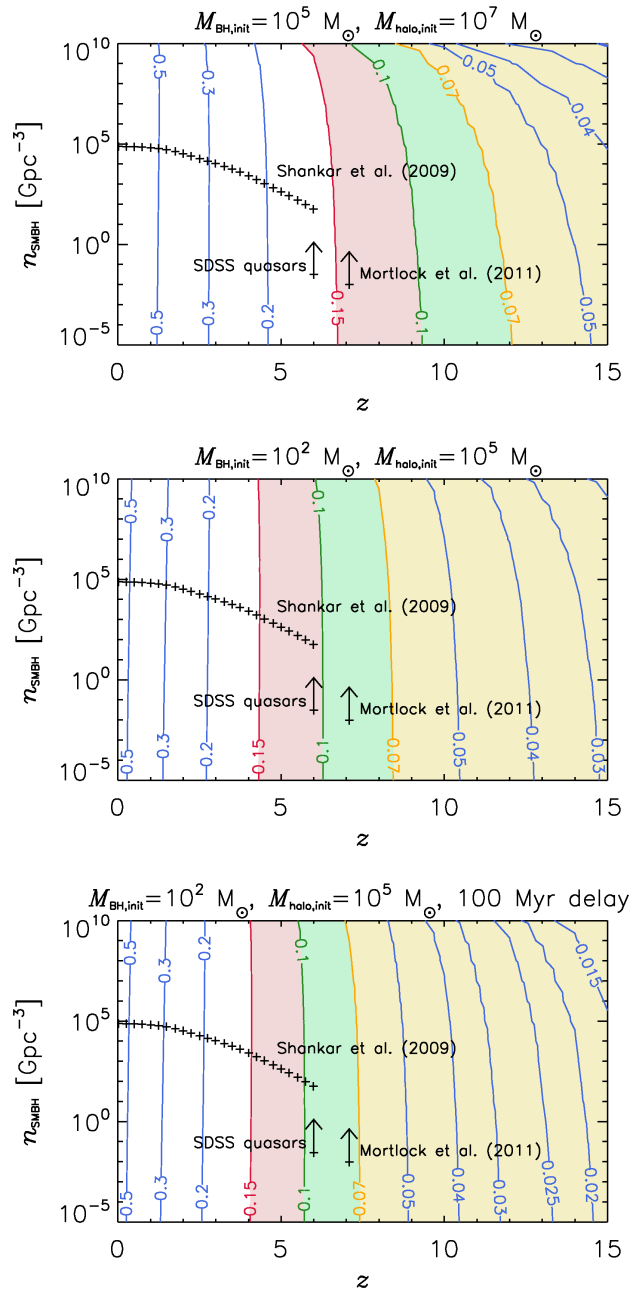}
       \caption{Upper limits for the radiative efficiency ($\epsilon$) required for various BH
         seeds to produce a space density $n_{\rm SMBH}$ (y-axes) of SMBHs with masses $M_{\rm
           BH,final}$$\ge$ 10$^9$ $M_{\odot}$ by redshift $z$ (x-axes),
         assuming accretion is limited to the Eddington rate ($f_{\rm
           Edd}$$f_{\rm duty}$ $\le$ 1).
         Contours show the maximum values of $\epsilon$ required, for three initial
         BH seeds:  $M_{\rm BH,init}$=10$^5$ $M_{\odot}$ SMS
         progenitors in $M_{\rm halo,init}$=10$^7$ $M_{\odot}$
         DM halos ({\it top}), 10$^2$ $M_{\odot}$ Pop~III progenitors in
         10$^5$ $M_{\odot}$ halos ({\it middle}), and
         the same case but with accretion delayed by 100 Myr due to
         radiative feedback ({\it bottom}).  The yellow, green and red contours correspond to
         $\epsilon$=0.07, 0.1 and 0.15, respectively; time-averaged super-Eddington growth is required in the
         corresponding colored regions, for these three cases, which
         are also shown in Fig. 2.
        The black crosses denote values of $n_{\rm SMBH}$ for SMBHs
        with masses $\ge$ 10$^9$ 
         $M_{\odot}$ inferred from observations.  A seed BH with $M_{\rm
           BH,init}$=10$^5$ $M_{\odot}$ ({\it top}) has to accrete continuously
         ($f_{\rm duty}$=1) at the Eddington rate ($f_{\rm Edd}$=1)
         with a radiative efficiency $\epsilon$ $\la$ 0.14 to explain
         the existence of the $z\simeq 7$ quasar, while a $M_{\rm
           BH,init}$=100 $M_{\odot}$ seed ({\it middle}) can only do so
         if accreting with a radiative efficiency $\epsilon$ $\la$ 0.09.
}
  \end{center}
\end{figure}

\section{Observational Constraints on Supermassive Black Hole Seeds}
We now turn to highlight the observational constraints on the 
accretion history of high-$z$ SMBHs that can be derived from the evolution of the
comoving number density of SMBHs over cosmic time.  In particular, we
place constraints on the time-averaged accretion rates and on the
efficiency with which radiation is produced during accretion, as
described below.

In general, the radiative luminosity, $L$, of a black hole accreting
at a rate $\dot{M}_{\rm BH}$ is given by $L = \epsilon\dot{M}_{\rm
  BH}c^2/(1-\epsilon)$, where $c$ is the speed of light and
$\epsilon$ is the radiative efficiency, defined as the fraction
of the rest mass energy of infalling matter that is converted into 
radiation during accretion.\footnote{We note that
  only a fraction (1-$\epsilon$) of the rest mass of the accreting
  material is finally accreted by the BH, as seen by an observer at
  infinity (e.g. Salpeter 1964; Thorne 1974).  The radiation generated
  during accretion which escapes to infinity accounts for the rest mass that such an observer 
  concludes is lost due to gravitational redshifting
  and time dilation as material falls into the potential well of
  the BH.}  We express the radiative luminosity as
a fraction $f_{\rm Edd}$ of the Eddington luminosity of the BH, $L$ = $f_{\rm
  Edd}$$L_{\rm Edd}$, where $L_{\rm Edd}$ = 1.2 $\times$ 10$^{38}$ erg s$^{-1}$ ($M_{\rm
  BH}$/$M_{\odot}$). The accretion rate at which a BH will radiate at a given 
fraction $f_{\rm Edd}$ of its Eddington luminosity $L_{\rm  Edd}$ is
then given by
\begin{equation}
\dot{M}_{\rm BH} = \frac{ (1-\epsilon) f_{\rm Edd} L_{\rm
    Edd}}{\epsilon c^2}  \mbox{\ .}
\end{equation}

We integrate equation (1) over time 
to find the final SMBH mass $M_{\rm BH,final}$, as a
function of its initial seed mass, $M_{\rm BH,init}$, the duty
cycle $f_{\rm duty}$ at which it accretes (defined as the fraction of
time spent accreting), and the radiative efficiency $\epsilon$.  
This yields 

\begin{eqnarray}
M_{\rm BH,final} & = & M_{\rm BH,init} \times \\ \nonumber
& &  {\rm exp}\left[\frac{f_{\rm Edd} f_{\rm
      duty}(1-\epsilon)}{\epsilon} \left(\frac{t_{\rm final}-t_{\rm  init}}{t_{\rm Edd}}  \right)\right] \mbox{\ .}
\end{eqnarray}
Here $t_{\rm Edd}$ = 450 Myr,\footnote{This timescale follows directly from the
definition of the Eddington luminosity: $t_{\rm Edd} = \sigma_{\rm T}
c / 4 \pi G m_{\rm p}$, where $\sigma_{\rm T}$ is the Thomson cross
section for electron scattering, $G$ is Newton's constant, $c$ is the
speed of light and $m_{\rm p}$ is the mass of the proton.} while $t_{\rm final}$ and $t_{\rm init}$
are the ages of the universe when the BH attains
its final mass and at the time of seed formation, respectively.

With $f_{\rm Edd}$ = 1 and $f_{\rm duty}$ = 1, equation (2)
reduces to the standard equation describing BH 
growth at the Eddington rate (defined as that which produces the Eddington
luminosity; e.g. Volonteri \& Rees 2006).  
The contours in Figure~1 denote the redshifts $z$ by which a number
density $n_{\rm SMBH}$ of black holes with masses $\ge$ 10$^9$ M$_{\odot}$ can
be grown, assuming constant accretion at the Eddington rate with radiative
efficiency $\epsilon$.  
For a given $\epsilon$, this allows us to constrain the growth histories
and initial seed masses of high-$z$ SMBHs from their observationally-derived number density.

For our comparison with the observational data in Fig.~1, we have expressed $t_{\rm final}$
as a function of redshift, using the {\it WMAP7}\/ results (Komatsu et
al. 2011) for the relevant cosmological parameters ($\Omega_{\rm M}$ =
0.27, $\Omega_{\rm \Lambda}$ = 0.73 and $H_{\rm 0}$ = 70.3 km s$^{-1}$
Mpc$^{-1}$), and following the approach in Barkana \& Loeb
(2001).\footnote{While we adopt the standard $\Lambda$CDM cosmological
model, the constraints on early BH growth are, of course, different in
different cosmological models (e.g. Melia 2013).} 
In order to determine $t_{\rm init}$ in equation (2), as a function of
$n_{\rm SMBH}$, we have used the Warren et al. (2006) DM halo mass function 
to find the highest redshift at which the comoving number density of the DM halos
with masses $\ge$ $M_{\rm halo,init}$ which can host BH seed formation
is $\ge$ $n_{\rm SMBH}$.  In principle, some of these halos could have
merged with one another before $t_{\rm final}$.  In this case, the
initial abundance of some of the seeds of SMBHs would have been
higher, suggesting that they formed at later times when the abundance
of their host halos was likewise higher.  Because, following our
arguments in Section 2, we neglect mergers in our
modeling, this implies that the values for $t_{\rm init}$ we adopt are lower limits. 

Finally, we note that the Warren et al. (2006) mass
function we have adopted provides a reasonable fit at the high redshifts
(e.g. $z$ $\simeq$ 30) of seed formation to the
halo mass functions found from cosmological simulations by Reed et
al. (2007).  That said,
Fig.~1 also shows that our results are relatively 
insensitive to the exact value of $n_{\rm SMBH}$, except at the
highest redshifts where there is the least time for SMBHs to
  form in large numbers (as shown by the downturn of the contours at
  the highest number densities).

In Fig. 1 we compare these theoretical curves with the observational
constraints on the evolution of the comoving number density $n_{\rm
  SMBH}$ of SMBHs over cosmic time.  
We show lower limits on $n_{\rm SMBH}$
for BHs with masses $\ge$ 10$^9$ $M_{\odot}$ at the highest redshifts,
as found from detections of $z$ $\simeq$ 6 quasars 
in the {\it Sloan Digital Sky Survey} (SDSS; e.g. Fan et al. 2006,
Tanaka \& Haiman 2009; see also Willott et al. 2010) and
from the single quasar at $z$ $\simeq$ 7 detected in the {\it United Kingdom 
Infrared Deep Sky Survey} (UKIDSS; Mortlock et al. 2011).\footnote{We
have estimated a conservative lower limit for $n_{\rm SMBH}$ at $z$
$\sim$ 7 by noting that just one $\ge$ 10$^9$ $M_{\odot}$ BH was found in the
UKIDSS Large Area Survey which covered $\simeq$ one tenth of the sky, and by
assuming that such a SMBH could have been detected out to $z$ = 10.
While this is a rough estimate, Fig.~1 shows that our results are
relatively insensitive to values of $n_{\rm SMBH}$ in this range.}
At lower redshifts, Fig.~1 also shows the number densities of $\ge$ 10$^9$
$M_{\odot}$ BHs obtained by integrating the BH mass function 
inferred from the observed luminosity function of active galactic
nuclei by Shankar et al. (2009).

In each of the three panels of Fig.~1 the
observed number densities of SMBHs are plotted together with 
the theoretical contours of $\epsilon$, 
for three different models of the initial BH seeds and their early
evolution.\footnote{For simplicity, we consider the two models described in Section~1 which bracket the
range of expected initial seed masses: the Pop~III remnant model and
the SMS remnant model.}  In the top panel, we take values of the initial seed 
mass $M_{\rm BH,init}$ = 10$^5$ $M_{\odot}$ and initial DM host halo mass $M_{\rm
  halo,init}$ = 10$^7$ $M_{\odot}$, corresponding to the SMS remnant model.  In the middle
panel, we take $M_{\rm BH,init}$ = 10$^2$ $M_{\odot}$ and $M_{\rm
  halo,init}$ = 10$^5$ $M_{\odot}$, corresponding to the Pop~III
remnant model.  Finally, in the bottom panel we take the same values
as in the middle panel but we assume accretion onto the seed BH is
delayed for the first 100 Myr; as
discussed in Section 4.2, numerous cosmological simulations
suggest that this may be the case for BHs born in minihalos at high-$z$ due to
radiative feedback from the progenitor star and from the accretion process
(e.g. Kitayama et al. 2004; Whalen et al. 2004; Yoshida 2006; Alvarez et al. 2009).

For a given value of $\epsilon$, the contours show the range of
redshifts over which there is sufficient time to grow a BH from its initial seed mass to
$M_{\rm BH,final}$ $\ge$ 10$^9$ $M_{\odot}$ assuming accretion at the
Eddington rate 
($f_{\rm Edd}$$f_{\rm duty}$ = 1).  In
particular, the yellow, green and red contours in Fig.~1
correspond to efficiencies $\epsilon$ = 0.07, 0.1 and 0.15, respectively.
The colored regions show the range of redshifts at which $\ge$ 10$^9$
$M_{\odot}$ BHs must have grown at time-averaged rates exceeding the
Eddington limit for each of these efficiencies. 
As these colored regions extend to lower
redshifts in the Pop~III remnant cases (bottom two panels), it is
clear that relatively low-mass Pop~III BHs could only be the seeds of
the SMBHs inferred at $z$ $\ga$ 6 in the SDSS and UKIDSS if
they grew at super-Eddington rates for a significant fraction of time
and/or if their radiative efficiency of accretion is relatively low,
$\epsilon$ $\la$ 0.09 (see e.g. Volonteri \& Rees 2006 for less
stringent constraints on $\epsilon$ gleaned from models based on a
higher, {\it WMAP1} $\sigma_{\rm 8}$ parameter
which implied much earlier seed BH formation).  However, the much more massive SMS remnant
seeds could accrete at sub-Eddington rates and still grow to $M_{\rm
  BH,final}$ $\ge$ 10$^9$
$M_{\odot}$ sufficiently rapidly, even allowing for a radiative
efficiency of up to $\epsilon$ $\simeq$ 0.14.

In the next two Sections we consider the values expected for the
parameters $\epsilon$, $f_{\rm Edd}$ and $f_{\rm duty}$, and what
they imply for models of SMBH seed formation.

\section{Suppression of Black Hole Growth via Local Radiative Feedback}
Here we discuss various ways in which radiation limits the rate of
growth of BHs in the early universe, and what values are expected for 
the radiative efficiency $\epsilon$ and the time-averaged accretion rates
$f_{\rm Edd}$$f_{\rm duty}$ of early BHs.  We then consider what these limits,
taken together, imply for the initial seed masses of SMBHs.

\subsection{Low Eddington Accretion Rate due to High Radiative Efficiency}
Following equation (1), for higher $\epsilon$ the Eddington luminosity is generated at
lower accretion rates, which implies that Eddington-limited accretion
proceeds at a lower rate for higher radiative efficiencies.  For
higher $\epsilon$ a BH must accrete at a higher
time-averaged accretion rate $f_{\rm Edd}$$f_{\rm duty}$ to grow to a
given mass within a given time.

Given that the Eddington accretion rate is sensitively dependent on
$\epsilon$ and that the radiative efficiencies of BHs can range from
$\epsilon$ $\simeq$ 0.025 to $\simeq$ 0.4 (see e.g. Milosavljevi{\' c} et al. 2009b
and references therein),\footnote{These values are expected for
  accretion through geometrically thin disks.  For spherically
  symmetric or advection dominated accretion (as expected at low accretion rates;
  e.g. Narayan \& Yi 1995), the radiative efficiency
can be much lower.} it is vital to constrain this
quantity in order to understand the growth of the earliest SMBHs 
(e.g. Shapiro 2005; King \& Pringle 2006; Volonteri \& Rees 2006; Tanaka et al. 2012).
Fortunately there is a growing body of observational evidence which
provides some guidance, and in particular suggests that SMBHs at
high-$z$ tend to have relatively high
radiative efficiencies, which suggests in turn that they are rapidly rotating.
Numerous authors have argued
that in order to account for the observations, SMBHs must have typical values 
of $\epsilon$ $\ga$ 0.1--0.15 (e.g. Elvis et al. 2002; Yu \& Tremaine 2002;
Volonteri et al. 2005; Shankar et al. 2010; but see e.g. Raimundo et al. 2012), and that $\epsilon$
tends to increase with redshift (Wang et al. 2006, 2009; Barausse
2012; Volonteri et
al. 2012; see also Maio et al. 2012) and with BH mass
(e.g. Davis \& Laor 2011; Shankar et al. 2011), with inferred values as high as
$\epsilon\sim0.3$--0.4 for $\ga10^9 M_{\odot}$ BHs at high-$z$ (Li et
al. 2012).  If the seeds of the highest-redshift SMBHs do indeed have such high
radiative efficiencies, then this would imply very strong constraints
on the initial masses of these seeds.  In particular, it would imply
much more massive seeds than can be explained in the Pop~III 
model, but which can be much more easily accommodated in the SMS model.

As shown in Fig.~1, for such high efficiencies (e.g. $\epsilon$ $\ga$ 0.15)
$\sim$100 $M_{\odot}$ Pop~III seed remnants would have to grow at super-Eddington
time-averaged accretion rates (i.e. with $f_{\rm Edd}$$f_{\rm duty}$
$>$ 1) in order to explain the Mortlock et
al. (2011) SMBH, the SMBHs powering the SDSS quasars and 
other $\ga$10$^9$ $M_{\odot}$ BHs at $z$ $\ga$ 4.  However,  more massive ($\ga$ 10$^5$ $M_{\odot}$) SMS remnants could grow
sufficiently fast to explain these high-$z$ SMBHs, even at
sub-Eddington rates. 

Fig.~1 also shows that for the highest
radiative efficiencies $\epsilon$ $\sim$ 0.3--0.4, even the supermassive seeds
from SMS
would have to accrete at super-Eddington time-averaged rates
(i.e. $f_{\rm Edd}$$f_{\rm duty}$ $>$ 1) just to grow to $M_{\rm
  BH,final}\ge 10^9\,M_{\odot}$ by $z\simeq2$--3.  This suggests that SMBHs with such
high radiative efficiencies at higher redshifts may have undergone
a period of rapid accretion with a lower radiative efficiency in
the past.  While we cannot know the entire accretion history of a
given SMBH at high-$z$, we can place some constraints on the amount of
high energy radiation emitted during its growth from the
observationally-inferred size of the H~{\sc ii} region surrounding
it.  For the case of the $z$ $\simeq$ 7 quasar, Mortlock et
al. (2011) report that it resides in an H~{\sc ii} region that is $\simeq$
1.9 physical Mpc in extent, which is smaller than those typically found
surrounding quasars at $z$ $\ga$ 6. This suggests that accretion may
indeed have been radiatively inefficient during much of the
growth of the BH.  Alternatively, however, the accreting BH could have
instead been obscured for much of its lifetime, resulting in a small fraction of ionizing
photons escaping into the intergalactic medium (see Bolton et
al. 2011).  This would be consistent with the galaxy merger-driven model
  for SMBH growth (e.g. Sanders et al. 1988; Hopkins et al. 2006) as well as with recent
observational evidence suggesting that a large fraction of accreting
SMBHs at high-$z$ are in fact buried within significant amounts of gas and
dust that prevent the escape of ionizing radiation (e.g. Fiore et
al. 2012; see also Kelly \& Shen 2012).\footnote{While Treister et
  al. (2011) argued for a large population of dust-obscured BHs at
  high-$z$, this result has been shown to be erroneous by Willott (2011).}

We emphasize that there are uncertainties in the estimated
radiative efficiencies that we have quoted here, in some cases of up
to a factor of a few (e.g. Li et al. 2012).  Nonetheless, the general
trends, as found from multiple measurements, that radiative
efficiencies increase with BH mass and with redshift are strongly
suggestive of high (e.g. $\epsilon$ $\ga$ 0.1) values for the most
massive BHs in the early universe, which are our focus in this work.

\subsection{Sub-Eddington Accretion due to Progenitor- and Accretion-Generated Radiation}
By definition, accretion is in principle limited to the Eddington
rate due to electron scattering of the photons generated in the
accretion process, but other radiative processes can further limit the
accretion rate.  Milosavljevi{\' c} et al. (2009a,b) find that
photoheating and pressure from ionizing and Lyman-$\alpha$
photons render the accretion of gas onto a BH intermittent ($f_{\rm
  duty}$ $<$ 1), with a time-averaged accretion rate of just $\sim$
0.3 times the Eddington rate (i.e. $f_{\rm Edd}$$f_{\rm duty}$ $\sim$
0.3). Park \& Ricotti (2011, 2012a,b) report similar results, although
they find that accretion at the Eddington rate can be achieved for
sufficiently dense accreting gas (see also Li 2011).
The results of larger-scale cosmological simulations 
also support the conclusion that accretion onto early BHs 
occurs at sub-Eddington rates due to accretion-generated radiative
feedback (e.g. Pelupessy et al. 2007; Alvarez et al. 2009; DeGraf et
al. 2012; Jeon et al. 2012).

An additional bottleneck to efficient accretion onto Pop~III seeds
born in minihalos is that 
the intense ionizing radiation emitted by their progenitor stars 
drives dense gas out of the halo (Kitayama et
al. 2004; Whalen et al. 2004; Alvarez et al. 2006), leaving the BH in a low-density medium
from which it cannot accrete rapidly (Yoshida 2006; Abel et al. 2007;
Johnson \& Bromm 2007).\footnote{Less massive (20--40 $M_{\odot}$)
  Pop~III progenitor stars may not
  evacuate the gas as completely, but the BHs they create are also likely to be ejected from
  their host halos due to kicks they receive during core collapse
  (Whalen \& Fryer 2012).}  This results in accretion rates orders of
magnitude below the Eddington rate for up to $\sim$ 10$^8$ years
before dense gas recollapses into the halo.  The bottom panel of
Fig.~1 shows the effect of such a delay in accretion onto a 100
$M_{\odot}$ Pop~III seed BH initially formed in a 10$^5$ $M_{\odot}$
DM halo.\footnote{We note that a similar delay in the formation of
  Pop~III seed BHs in low-mass DM halos could also result from
  the super-sonic streaming of dark matter halos relative to the gas,
  as discussed by e.g. Greif et al. (2011), Maio et al. (2011) and
  Stacy et al. (2011).}  Comparing this to the case without a delay (middle
panel) shows that the effect is comparable to a decrease of 
$\sim 10$--20\% in the average accretion rate onto the seeds of SMBHs formed 
by $z\sim 6$--8.  Thus, the radiative feedback from Pop~III
progenitor stars can significantly slow down the growth of
Pop~III seeds to SMBHs.  

While radiative feedback from both accretion and the
progenitor star are likely to limit $\dot{M}_{\rm BH}$ to
sub-Eddington values at early times, we note that 
SMBH-powered quasars at $z$ $\sim$ 6 are inferred to have Eddington
ratios ($f_{\rm Edd}$) near unity.  Their duty cycles ($f_{\rm duty}$),
however, are not well-constrained (Willott et al. 2010; see also Shankar et al. 2010),
although they are likely to be larger than those inferred for SMBHs at lower
redshifts (e.g. Trakhtenbrot et al. 2011).
Overall, however, we conclude that radiative feedback, at least at the
earliest times, appears likely to keep the time-averaged accretion rate of
Pop~III remnant seeds at $f_{\rm Edd}$$f_{\rm duty}$ $<$ 1.

Accretion-generated radiative feedback may also limit the growth of 
SMS remnant BH seeds to sub-Eddington rates (e.g. Johnson et
al. 2011).  However, as the intense ionizing
radiation emitted by rapidly growing SMSs cannot escape their heavy
accretion flows, their host halos are less likely to be photoevaporated (Johnson et
al. 2012b; see also Hosokawa et al. 2012); in turn, 
the BH remnants they leave behind are likely to accrete more rapidly
than Pop~III remnant seeds.  We note that this is in basic agreement 
with the suggestion by Salvaterra et al. (2012) that more massive 
seed BHs accrete with higher Eddington fractions than do lower mass seeds.

\section{Implications and Discussion}
In the previous Sections we have highlighted the constraints
that observations of high-$z$ quasars place on the nature of the BH seeds and
subsequent growth of SMBHs in the early universe, and we have reviewed
the ways in which radiative feedback from stars and accreting BHs is
expected to limit the growth of these objects.  We now discuss the implications
of these findings for the main models of SMBH seed formation.

In Sections 2 and 3 we showed that mergers of BH seeds are likely to play only a
minor role in growing SMBHs, and that the rate of accretion required to grow the observed
SMBHs depends only weakly on their number density (see Fig.~1).
Thus, we are justified in taking the simplified approach of solving equation (2)
for the initial seed mass $M_{\rm BH,init}$ as a function of
$\epsilon$, $f_{\rm Edd}$, and $f_{\rm duty}$, without regard for the
role of mergers or for the precise redshift of formation of the seeds.  Figure 2 shows the
minimum BH seed mass, $M_{\rm BH,init}$, required to grow a SMBH
to a mass of $10^9\,M_{\odot}$ by redshift $z$, with the
three panels corresponding to the scenarios shown in the panels in
Fig.~1.  In each panel, the minimum BH seed mass is shown for two
choices of the time-averaged Eddington fraction ($f_{\rm Edd}$$f_{\rm
  duty}$ = 0.5 and 1) and for the same three radiative efficiencies
highlighted in Fig.~1 ($\epsilon$ = 0.07, 0.1, and 0.15).  For each of
these cases we have assumed a value of $t_{\rm init}$ corresponding to
a space density of SMBHs of $n_{\rm 
  SMBH} = 1\,\mbox{Gpc}^{-3}$ (comoving), but as Fig.~1 shows the results
are not strongly sensitive to this choice.  

\begin{figure}
  \begin{center}
    \leavevmode
      \epsfxsize=8.6cm\epsfbox{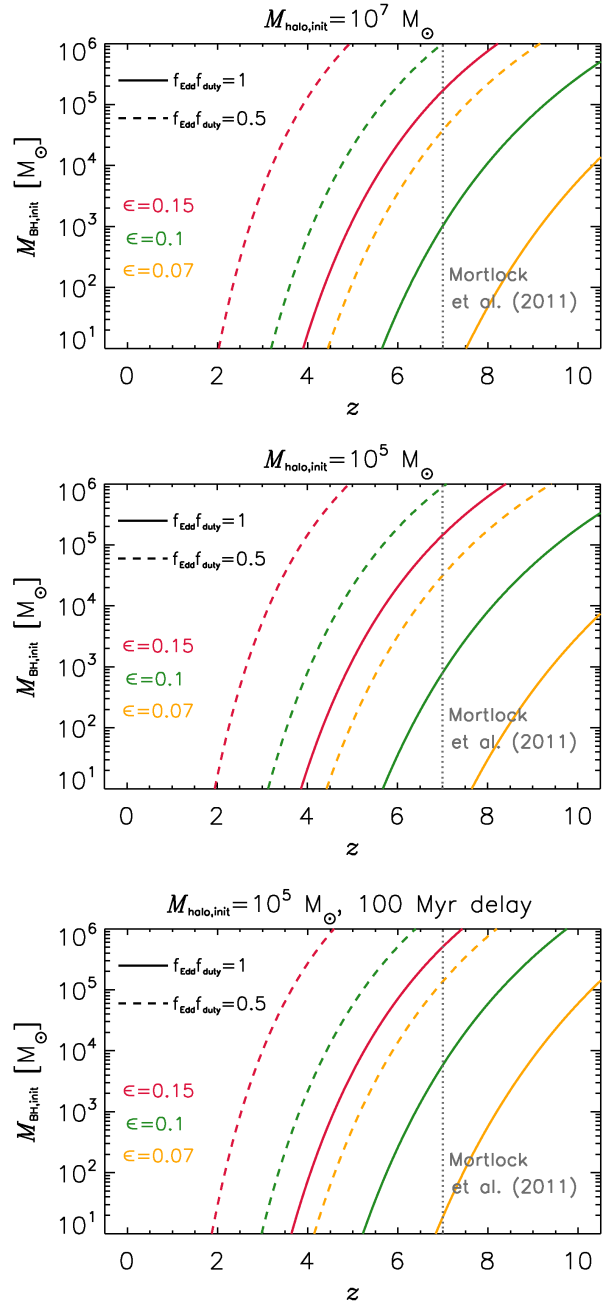}
       \caption{Minimum initial BH seed mass M$_{\rm BH,init}$
       required to form a $M_{\rm BH,final}$ $\ge$ 10$^9$ $M_{\odot}$ SMBH by redshift $z$,
       at time-averaged Eddington fractions of $f_{\rm Edd}$$f_{\rm
         duty}$ = 1 ({\it solid
         lines}) and 0.5 ({\it dashed lines}),
       for three different radiative efficiencies:
       $\epsilon$ = 0.15 ({\it red}), 0.1 ({\it green}), and 0.07 ({\it
         yellow}).  The three panels correspond to the cases 
       shown in the panels of Fig.~1.  
         The largest seed masses are
       implied for the case of seed formation in a $M_{\rm halo,
         init}$=10$^5$ $M_{\odot}$
       halo and accretion delayed by 100 Myr due to radiative feedback
       ({\it bottom panel}), while less stringent constraints are given
       for the cases with no delay ({\it
         middle panel}) and with
       larger initial halo mass ($M_{\rm halo,init} = 10^7\,M_{\odot}$ instead
       of $M_{\rm halo,init} = 10^5\,M_{\odot}$) ({\it top panel}).  Given the high radiative
       efficiencies ($\epsilon$ $\ga$ 0.1--0.15) inferred for high-$z$ SMBHs, these objects likely
       grew via constant (or time-averaged) super-Eddington
       accretion and/or started as massive $M_{\rm BH,init}$$\ga$
       10$^3$--10$^5$ $M_{\odot}$ BH seeds. The dotted gray vertical
       line denotes the redshift of the Mortlock et al. (2011)
       quasar.  Compared to the earlier 10$^9$ M$_{\odot}$ SMBHs
       uncovered in the SDSS at $z$ $\simeq$ 6, this single quasar
       implies a seed mass that is an order of magnitude higher.}
  \end{center}
\end{figure}

The top two panels differ by very little, which is a reflection of the
fact that the time between the formation of the first
10$^5$ $M_{\odot}$ halo and that of the first 10$^7$ $M_{\odot}$ halo
in a 1 Gpc$^3$ comoving cosmological volume is small compared to the
time from their formation to redshift $z$ (on the horizontal axis). 
The values of $M_{\rm BH,init}$ shown in the bottom panel,
however, are somewhat higher, reflecting the fact that the 100 Myr
delay assumed in this case is longer than the time between the
formation of the first 10$^5$ and 10$^7$ $M_{\odot}$ halos.   In every
case, it is clear that high radiative efficiencies ($\epsilon\ga 0.1$--0.15),
such as those inferred from observations of high-$z$ SMBHs 
(see Section 4.1), imply seed masses for observed SMBHs at $z\ga 7$
of $M_{\rm BH,init}$ $\ga$ 10$^3$--10$^5$
$M_{\odot}$, assuming constant Eddington-limited accretion ($f_{\rm
  Edd}$$f_{\rm duty}$ = 1).  For lower time-averaged accretion   
rates, as suggested by simulations of BH accretion (see Section 4.2), 
the implied minimum seed masses are much higher, up to $M_{\rm BH,init}\ga
10^6\,M_{\odot}$ for $f_{\rm Edd}f_{\rm duty}\sim0.5$ 
over the same range of radiative efficiencies.

If accretion is radiatively efficient and the
time-averaged accretion rate is sub-Eddington, as discussed in Section~4, then
the initial seed masses of the highest redshift SMBHs must
have been very high, perhaps exceeding even the range predicted by
the SMS remnant model (10$^4$--10$^6$ $M_{\odot}$) in order to
explain the 2 $\times$ 10$^9$ $M_{\odot}$ BH reported at $z$ $\simeq$
7 (Mortlock et al. 2011).  Interestingly, this is also consistent with
the results of recent large-scale
cosmological simulations tracking SMBH growth which suggest that such
high-$z$ quasars can be explained by starting with initial seed BH masses
of 10$^5$ $M_{\odot}$ (Li et al. 2007; Di Matteo et al. 2012).  Taken together, we conclude that 
the available theoretical and observational evidence
strongly suggests that the seeds of SMBHs at high-$z$ were likely very massive.  In turn, as the SMS remnant
model produces the most massive seeds of the three main models
discussed in the introduction, the evidence suggests that this model
may be the most viable (see also e.g. Natarajan \& Volonteri
2012).\footnote{Our generic conclusion is that higher initial seed
  masses are more consistent with the available observational data and
  theoretical modeling.  In this, we conclude that the ($\sim$ 100
  $M_{\odot}$) Pop~III seed model is the weakest, the ($\sim$
  10$^3$ $M_{\odot}$) stellar cluster seed model is somewhat 
  more favorable, and the (10$^4$--10$^6$ $M_{\odot}$) SMS seed
  model is the strongest.}

One of the arguments against the SMS
remnant model has been that SMSs are exotic objects, and they
may never form in our universe.  A growing body
of work is elucidating the conditions required for these objects to form
(e.g. Bromm \& Loeb 2003; Koushiappas et al. 2004; Lodato \&
Natarajan 2006; Begelman et al. 2006; Spaans \& Silk 2006; Regan \& Haehnelt 2009; Shang et
al. 2010; Schleicher et al. 2010; Ball et al. 2011; Wolcott-Green et al. 2011; see also Mayer et al. 2010
on massive seed formation from metal-enriched gas),\footnote{We note
  that while Mayer et al. (2010) argue for this alternative route to
  SMS formation, they require that these objects form in $\sim$
  10$^{11}$ M$_{\odot}$ halos, which would form much later than the 10$^7$
M$_{\odot}$ halos we have focused on here. As this leaves far less time for
their growth to 10$^9$ M$_{\odot}$, it is unlikely that these can be
the seeds of the highest-$z$ SMBHs.}  and there are suggestions
that the conditions for SMS formation  may occur much more often in the early universe than previously
assumed (Dijkstra et al. 2008; Agarwal et al. 2012; Johnson et
al. 2012a; Petri et al. 2012).  Indeed, Agarwal et al. (2012) find
that, due to locally high LW fluxes generated by Pop~II stars in the
early universe,
a sufficient number of SMSs may form to provide the seeds for a large
fraction of the SMBHs in the centers of galaxies today (see also
Greene 2012).  This development
offers a completely independent reason to seriously consider SMS remnants. 

While the SMS remnant model offers an explanation for the origins of
the earliest SMBHs that is consistent with the available data, we
note that other models starting with lower initial seed masses
can, in principle, also explain the data.   For instance, as shown in
Fig.~2, if $M_{\rm BH,init}$ $\sim$ 10$^2$ $M_{\odot}$ seeds accrete gas with very low radiative
efficiency ($\epsilon$ $\la$ 0.09) continuously at the Eddington rate
or, perhaps intermittently, above it (e.g. Jaroszynski et al. 1980; Collin et al. 2002;
Kawaguchi et al. 2004; Ohsuga et al. 2005; Volonteri \& Rees 2005; 
Kurosawa \& Proga 2009; Wyithe \& Loeb 2011; Begelman 2012a; Li 2012)
then, even allowing for a $\sim$ 100 Myr delay due to radiative
feedback (bottom panel), they could grow to $\ga$ 10$^9$ $M_{\odot}$
by $z$ $\ga$ 7.\footnote{We also note that Umemura et al. (2012) have argued
that Pop~III seeds could form in much smaller $\sim$ 10$^4$
M$_{\odot}$ halos, which would have formed at earlier cosmological
times.  If this result is confirmed, it would suggest somewhat 
reduced constraints on the Pop~III seed model for SMBH formation.}  
However, it is possible that a large fraction of the
mass in such super-Eddington flows is lost to a wind instead of being
accreted onto the BH (see e.g. Begelman 2012b; Dotan \& Shaviv 2011).
It also remains to be demonstrated that these
conditions are likely to be realized at $z$ $\ga$ 7.  To the contrary, the available observational
evidence suggests high radiative efficiencies (e.g. $\epsilon$ 0.1--0.15 or higher; see
Section 4.1) and the available theoretical modeling suggests sub-Eddington 
accretion (see Section 4.2); as we have discussed here, the SMS seed model can explain 
the presence of the highest-$z$ SMBHs, even given such high radiative efficiencies and limited accretion rates,
while the Pop~III seed model cannot. 

It is difficult to verify the SMS remnant model without direct observational evidence of the existence of SMSs.
We note, however, that there are observational signatures of SMSs that
may be detected by future missions such as the {\it James Webb Space
  Telescope} (e.g. Johnson et al. 2012b).  They may also leave
unique chemical signatures that could be detected in Lyman-limit
systems (Woosley 1977; Fuller \& Shi 1997).  In addition, upon their
collapse SMSs are predicted to emit a large neutrino flux that could
be detectable (by e.g. {\it IceCube}) (Shi \& Fuller 1998; Linke et al. 2001; Fryer \& Heger 2011;
Montero et al. 2012), as well as to produce gravitational wave signatures that could be uncovered
by the {\it Laser Interferometer Gravitational Wave Observatory}~(e.g. Fryer \& New
2011),\footnote{See also e.g. Barausse (2012) on
  distinguishing between the SMS and Pop III remnant models using the
  gravitational wave signal of BH mergers.} and extremely
bright supernovae (e.g. Fuller et al. 1986) that could be found by the {\it Wide-Field Infrared
  Survey Telescope} (e.g. Whalen et al. 2012).
In lieu of such observations, the best evidence
for the existence of SMSs remains the SMBHs to which their remnants may
have grown at the highest redshifts.

\section{Summary}
 In closing, we provide a summary of our new results and conclusions:

\begin{itemize}

\item With proper accounting for the masses of the halos in which Pop~III
  stars form at high redshift, as well as for the suppression of the
  Pop~III star formation rate due to the build-up of the LW
  background radiation field, we conclude that mergers played a limited role in
  the growth of Pop~III seed BHs (see Section 2).

\item Because the time available for the growth of seed BHs to the
  $\ga$ 10$^9$ $M_{\odot}$ SMBHs observed at high redshift is only weakly
  dependent on the number density $n_{\rm SMBH}$ of these objects, we
  can safely assume that all such SMBHs have roughly the same amount
  of time to grow by a given redshift (see Section 3).  This allows us to estimate the
  minimum initial seed mass required to grow them, as a function of
  just the radiative efficiency $\epsilon$ and the time-averaged
  fraction of the Eddington rate at which they accrete, $f_{\rm
    Edd}$$f_{\rm duty}$.

\item Using the most recent cosmological parameters (from {\it
    WMAP7}), we have shown that the highest-redshift SMBHs known can
  only be explained by the Eddington-limited growth of seed BHs with
  masses of $M_{\rm BH,init}$ $\sim$ 100 $M_{\odot}$ (Pop~III seeds) and $\sim$ 10$^5$ $M_{\odot}$ (SMS seeds), if the radiative efficiency of
  accretion is $\epsilon$ $\la$ 0.09 and $\la$ 0.14, respectively (see Section
  3).  Accounting for the likely
  suppression of accretion at early times, due to radiative feedback
  from the BH seed progenitor stars and from accretion onto the seeds
  themselves, leads to even tighter
  constraints on the Pop~III seed model (see Section 4.2).

\item In turn, the high radiative efficiencies that are estimated for the highest-redshift and most
  massive SMBHs ($\epsilon$ $\ga$ 0.1--0.15; see Section 4.1) are much more easily
  accommodated in the SMS seed model than in the Pop~III seed model, given the much larger initial seed masses expected in
  the former (see Section 5).  This is especially true if the
  time-averaged accretion rates of seed BHs are sub-Eddington, as 
  suggested by much recent theoretical work (see Section 4.2).

\end{itemize}

\section*{Acknowledgements}
This work was supported by the U.S. Department of Energy
through the LANL/LDRD Program, and JLJ acknowledges the 
support of a LDRD Director's Postdoctoral Fellowship at Los Alamos
National Laboratory.  DJW acknowledges support from the Bruce and
Astrid McWilliams Center for Cosmology at Carnegie Mellon University.
DEH acknowledges support from National Science Foundation CAREER grant PHY-1151836.
The authors thank Jennifer Donley, Xiaohui Fan, Chris Fryer and Marta Volonteri 
for valuable feedback on early drafts of this work, Brian O'Shea for
kindly providing the code used to compute the Warren mass functions, and J.~J. Cherry,
Stirling Colgate, Dave Collins, George Fuller,
Xiaoyue Guan, Joe Smidt, Mike Warren, and Hao Xu for helpful
discussions.  This work benefited from the comments of anonymous reviewers.

\bibliographystyle{apj}

\end{document}